\documentstyle[12pt]{article}
\begin{document}
\newcommand{\kslash}{\mbox{$\displaystyle\not\mkern-4mu k$}}
\newcommand{\Dslash}{\mbox{$\displaystyle\not\mkern-4mu D$}}
\newcommand{\be}{\begin{equation}}
\newcommand{\ee}{\end{equation}}
\mbox{ }\hfill{\normalsize ITP-93-33E}\\
\mbox{ }\hfill{\normalsize May 1993}\\
\begin{center}
{\Large \bf High-$Q^{2}$  Elastic $ed$-scattering and\\[.5cm]
QCD Predictions}\\
\vspace{1cm}
{\large A.~Kobushkin and A.~Syamtomov}\\[.5cm]
{\large \it N.N.Bogolyubov Institute for Theoretical Physics, \\
Ukrainian Academy of Sciences, Kiev 143, Ukraine
}
\end{center}
\date{}
\vspace{.5cm}
\begin{abstract}
In the framework of pertubative QCD it is argued that in the elastic
$ed$-scattering at $Q^{2}\sim$ few $(GeV/c)^{2}$ the light-cone-frame
helicity-flip amplitudes could not be omitted. The obtained $\frac{B}{A}$ ratio
of Rosenbluth structure functions is shown to be in a good agreement with
experimental data. The high $Q^{2}$ behavior of $T_{20}$ is discussed.
\end{abstract}
\newpage
In a recent paper, Brodsky and Hiller \cite{Brod} have made a general
analysis of the e.m. properties of spin-one systems prceeding from the
gauge theory. They have mentioned  that in the analysis of high energy
elastic $ed$-cross-sections at least two momentum scales must be
distinguished. The first one is given by the QCD scale \mbox
{$\Lambda_{QCD}\approx200MeV/c$} and determines the perturbative
QCD regime. In particular, according to arguments of Ref.\cite{Gross},
this scale controls a high $Q^{2}$ suppression of the light-cone-frame
helicity-flip matrix elements of the deuteron e.m. current

\be
\begin{array}{rcl}
G_{+0}^{+}\approx{
a\left(\frac{\displaystyle{{\Lambda}_{QCD}}}{\displaystyle{Q}}\right)
G_{00}^{+}},\\
G_{+-}^{+}\approx{
b\left({\frac{\displaystyle{{\Lambda}_{QCD}}}{\displaystyle{Q}}}
\right)^{2}
G_{00}^{+}},
\label{Term1}
\end{array}
\ee
where
$G_{h^{'}h}^{+}(Q^{2})\equiv{<p^{'}h^{'}\mid{J^{+}}\mid{ph>}}$ are the
matrix elements of the e.m. current plus component
\mbox{$J^{+}\equiv{J^{0}+J^{3}}$}; $\mid{ph>}$ is an eigenstate of
momentum $p$ and helicity $h$; $a$ and $b$ are some constants.

 The second scale is given by the deuteron mass $M$ and has a purely
kinematic origin. In Ref.\cite{Brod} it was argued that in the
region

\be
Q^{2}{\gg}2M{\Lambda}_{QCD}\approx 0.8(GeV/c)^{2}
\label{Term2}
\ee
helicity-conserving matrix element $G_{00}^{+}$ dominates. For the
dimensionless ratio \mbox{${\eta}=\frac{\displaystyle{Q^{2}}}
{\displaystyle{4M^{2}}}$} (which appears in expressions for the
deuteron e.m. form factors given by $(3.7)$ of Ref.\cite{Brod}) this
condition translates into \mbox{$\eta{\gg}\frac{\displaystyle{
{\Lambda}_{QCD}}}{\displaystyle{2M}}\approx0.05$}. Omitting the
helicity-flip matrix elements $G_{+0}^{+}$ and $G_{+-}^{+}$, the
following  ratios for the charge, magnetic and quadrupole form factors
were derived \cite{Brod}
\be
G_{C}(Q^{2}):G_{M}(Q^{2}):G_{Q}(Q^{2})=\left(1-\frac{2}{3}\eta
\right):2:-1.
\label{Term3}
\ee
One can  note here that the asymptotical relation
$G_{C}\approx\left(\frac{\displaystyle{Q^{2}}}{\displaystyle
{6M^{2}}}\right)G_{Q}$ of \mbox{Refs.\cite{Gross,Car}} is reproduced from
(\ref{Term3}) at considerably high momentum transfer \mbox{$Q^{2}
\gg20(GeV/c)^{2}$}.

 In the present paper we would like to mention that the helicity-flip
matrix element $G_{+0}^{+}$ cannot be neglected. For example, in the
light-cone frame defined in Ref.\cite{Brod} the deuteron magnetic form
factor is
\be
G_{M}=\frac{1}{p^{+}(2\eta+1)}\left[(2\eta-1)\frac{G_{+0}^{+}}
{\sqrt{2\eta}}+G_{00}^{+}-G_{+-}^{+}\right],
\label{Term4}
\ee
where $p^{+}$ is a plus component of the deuteron momentum.
According to (\ref{Term1}) the contribution from $G_{+-}^{+}$ is
suppressed by the factor $\left(\frac{\displaystyle{\Lambda_{QCD}}}
{\displaystyle{Q}}\right)^{2}$. But the contribution from $G_{+0}^{+}$
is {\em{enhanced}} by the kinematical factor $-\sqrt{\frac
{\displaystyle{1}}{\displaystyle{2\eta}}}$ at $\eta\ll\frac{\displaystyle1}
{\displaystyle2}$
and by the factor $\sqrt{2\eta}$ at \mbox{$\eta\gg\frac{\displaystyle1}
{\displaystyle2}$}.
It must be also stressed that the signs from $G_{+0}^{+}$
contributions are different for the small $\eta$-value and large
$\eta$-value regions. So there may be a point \mbox{$Q^{2}=Q_{0}^{2}$}
where the contribution from $G_{+0}^{+}$ conceals $G_{00}^{+}$  matrix
element and the magnetic form factor changes its sign. For deuteron
this occurs approximately at $2(GeV/c)^{2}$, what
experimental data tell us about \cite{Arnold}. In this case the
coefficient $a$ in the expression (\ref{Term1}) must be equal to

\be
a={\frac{\sqrt{2}Q_{0}^{2}M}{(2M^{2}-Q_{0}^{2})\Lambda_{QCD}}}
\approx 5.08.
\label{Term5}
\ee
Adopting this assumption, we obtain the following expressions for
the deuteron electromagnetic form factors in the  region under
consideration
\begin{eqnarray}
G_{C}=\frac{1}{2p^{+}(2\eta+1)}\left(\frac{6M^{2}+5Q_{0}^{2}}
{6M^{2}-3Q_{0}^{2}}-\frac{2}{3}\eta\right)G_{00}^{+},\nonumber\\
G_{M}=\frac{1}{p^{+}(2\eta+1)}\left(\frac{2\eta-1}{2\eta}
\frac{Q_{0}^{2}}{2M^{2}-Q_{0}^{2}}+1\right)G_{00}^{+},\\
G_{Q}=\frac{1}{2p^{+}(2\eta+1)}\left(\frac{1}{\eta}\frac{Q_{0}^{2}}
{2M^{2}-Q_{0}^{2}}-1\right)G_{00}^{+},\nonumber
\label{Term6}
\end{eqnarray}
and, therefore, the ratio of the Rosenbluth structure functions
${B}\over{A}$ and tensor analysing power $T_{20}$ become $G_{00}^{+}$
-independent.The results of numerical calculations for ${B}\over{A}$
and $T_{20}$ in comparison with experimental data and the results of
Ref.\cite{Brod} are shown in Figs. 1 and 2,
respectively, where $Q_{0}^{2}$ parameter is chosen to be
$1.93(GeV/c)^{2}$.

 At very high $Q^{2}$ (when $\eta\sim$ 1) the ratios (\ref{Term3})
are modified to:
$$
G_{C}(Q^{2}):G_{M}(Q^{2}):G_{Q}(Q^{2})  =
\left(\frac{2}{3}\eta-
\frac{6M^{2}+5Q_{0}^{2}}{6M^{2}-3Q_{0}^{2}}\right):
$$
\be
  2\left(\frac{1-2\eta}{2\eta}\frac{Q_{0}^{2}}{2M^{2}-
Q_{0}^{2}}-1\right):\left(\frac{1}{\eta}\frac{Q_{0}^{2}}{2M^{2}-
Q_{0}^{2}}-1\right).
\label{Term7}
\ee
Omitting helicity-flip matrix elements, Brodsky and Hiller
{\cite{Brod}} showed that at large $Q^{2}$ the ratios of deuteron
e.m. form factors are identical to those of the point-like spin-one
fields. From (\ref{Term7}) one sees that such statement remains valid
only for \mbox{$G_{C}:G_{Q}$} and only when $\eta\gg1$.

 Some comments should be made about the gross structure of the deuteron
e.m. form factors. As was already mentioned, experiment shows that at
$Q^{2}=Q_{0}^{2}$ the magnetic form factor
changes its sign. This means that it is negative for \mbox{$Q^{2}>
Q_{0}^{2}$} and is positive for \mbox{$Q^{2}<Q_{0}^{2}$}. Therefore
the matrix element $G_{00}^{+}$ (according to (6)) should be
{\em{negative}}. It means (see ratios (\ref{Term7})) that $G_{C}$ is
negative within the interval from \mbox{$Q^{2}\sim0.8(GeV/c)^{2}$}
(the margin, where the perturbative QCD regime becomes to be open) to
\mbox{$Q^{2}\approx11M^{2}$}.  Since $G_{C}$ is positive at very
small $Q^{2}$ one can make a conclusion that $G_{C}$ has to change its
sign at $Q\sim$ few hundred MeV/c. Such structure is in a good
agreement with conventional meson+nucleon picture predictions (see,
e.g. \cite{Gari}), which are assumed to be valid for \mbox{$Q^{2}\leq 1
(GeV/c)^{2}$}. As for the deuteron quadrupole form factor, it is
negative in the region where the perturbative QCD is valid, but

\be
Q^{2}\leq \frac{4M^{2}Q_{0}^{2}}{2M^{2}-Q_{0}^{2}}\approx5.32
(GeV/c)^{2}.
\ee
On the other hand at the origin it is positive. This means that
contrary to the meson-nucleon picture, $G_{Q}$ has a node at some
value $Q^{2}$, where perturbative QCD is not yet applicable.

    Concluding we should emphasize that our approach is not in
contradiction with the $G_{00}^{+}$ dominance hypothesis, which,
in our opinion, together with available experimental data may allow
us to make a reasonable predictions for deuteron electromagnetic form
factor behavior at large momentum transfer. We plan to develop an
elaborate analysis of one of possible scenarios of such evaluation in
a more detailed paper.\\

The authors are grateful to S. Brodsky for encouragement of this work
and to J.~Hiller for sending $B/A$ experimental data.
\newpage
\begin{center}
Figure Captions\\
\end{center}
\vspace{0.1cm}
\begin{description}
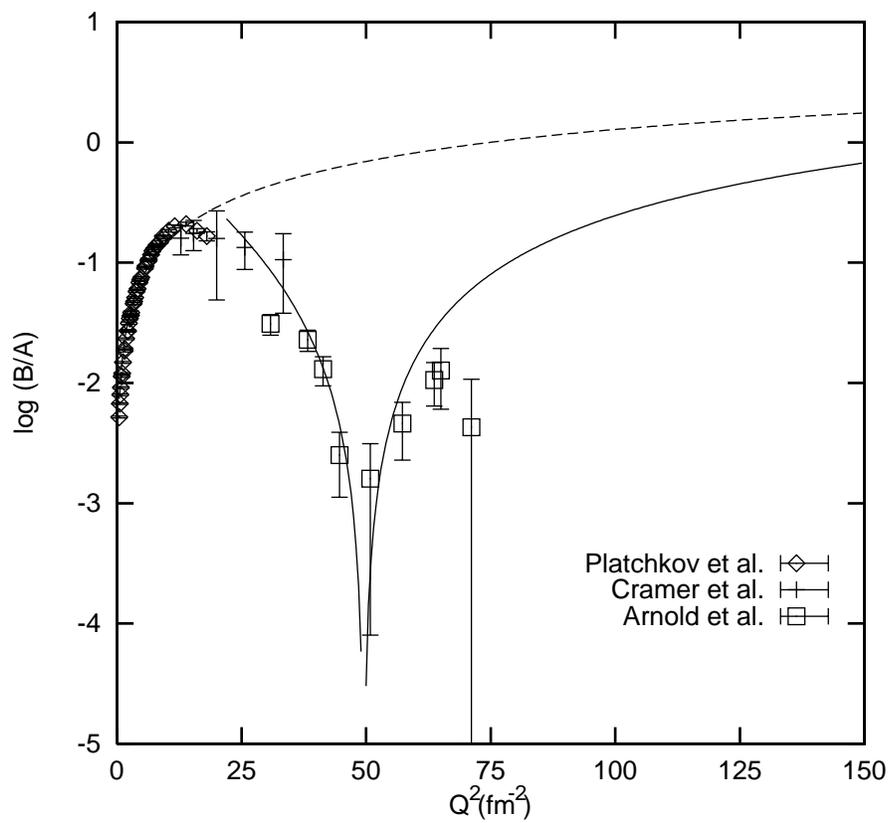
\item[Figure 1.] Comparison of our calculations (solid line) for $B/A$
ratio with experimental data Ref.\cite{Arnold}. The results of
calculations of Ref.\cite{Brod} are plotted by dashed line.
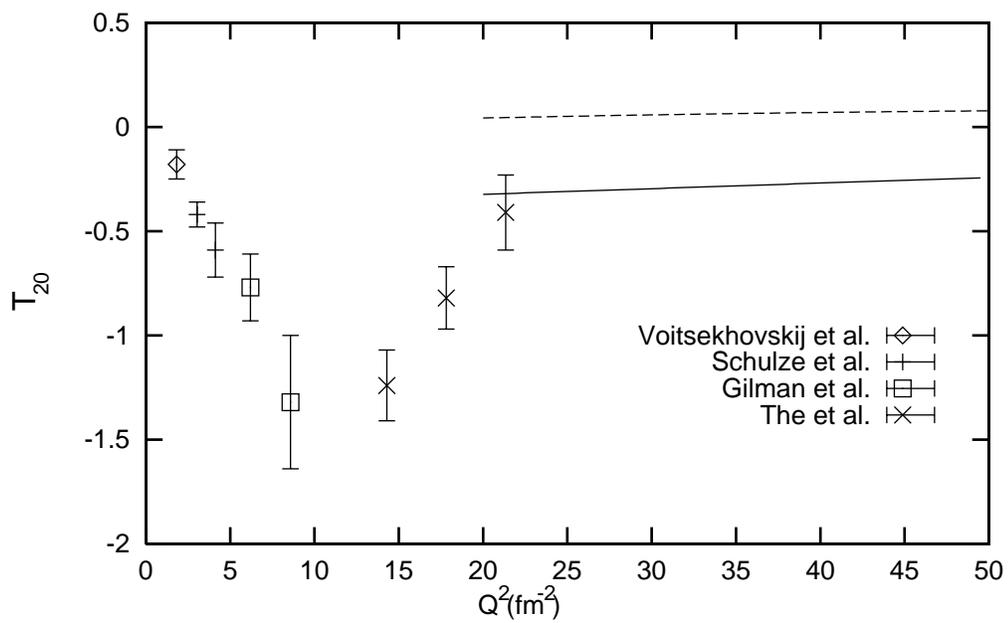
\item[Figure 2.] The predictions for $T_{20}$ in our approach (solid line)
and the model of Ref.\cite{Brod} (dashed line) at $\theta=0^{0}$.
Experimental data are from Refs.\cite{The}.
\end{description}

\end{document}